\documentclass[fleqn, usenatbib]{mnras}

\usepackage{graphicx}	
\usepackage{amsmath}	
\usepackage{amssymb}	
\usepackage{multicol}   
\usepackage[T1]{fontenc}
\usepackage{newtxtext, newtxmath}

\title[Early-type Dwarf Galaxies in the Local Universe]{Early-type Dwarf Galaxies in the Local Universe. Evidence of Ex-situ Growth}

\author[Paudel et al.]{
Sanjaya Paudel,$^{1,2}$
Suk-Jin Yoon,$^{1,2}$\thanks{E-mail:sjyoon0691@yonsei.ac.kr }
Jun-Sung Moon,$^{1,3}$
Daya Nidhi Chhatkuli$^{4}$
\\
$^{1}$Department of Astronomy, Yonsei University, Seoul, 03722, Republic of Korea\\
$^{2}$Center for Galaxy Evolution Research, Yonsei University, Seoul, 03722, Republic of Korea\\
$^{3}$Astronomy Program, Department of Physics and Astronomy, Seoul National University, Republic of Korea\\
$^{4}$Central Department of Physics, Tribhuvan University, Kirtipur, Kathmandu, Nepal
}

\date{Accepted \today. Received \today.; in original form \today.}

\pubyear{2022}

\begin{document}
\label{firstpage}
\pagerange{\pageref{firstpage}--\pageref{lastpage}}
\maketitle

\begin{abstract}
We report the discovery of a rare early-type dwarf galaxy (dE), SDSS\,J125651.47+163024.2 (hereafter dE1256), possessing a tidal feature that  was likely built up by accretion of an even smaller dwarf galaxy. 
dE1256 is located in a nearly isolated environment, at the outskirt of the Virgo cluster. 
A detailed morphological examination reveals that the accreted stellar population is mainly deposited in the outer part of dE1256, where the tidal tail is most prominent.
The inner part of dE1256 is perfectly modeled with a simple S\'ersic function of index n = 0.63 and half-light radius R$_{h}$ = 0.6 kpc, but in contrast, the entire galaxy has a size of R$_{h}$ = 1.2 kpc. 
The mass ratio between the host and the putative accreted dwarf galaxy is calculated to be 5:1, assuming that the observed two components, inner S\'ersic and outer tidal tail residual, represent the host's and accreted galaxy's stellar populations, respectively. We suggest that while the accretion contributes only 20\% of the overall stellar population, the size of dE1256 grew by a factor of two via the accretion event.
Our results provide, for the first time, strong observational evidence that a dE is undergoing a two-phase growth, a common phenomenon for massive galaxies.
\end{abstract}


\section{Introduction}

The size of massive early-type galaxies (Es) has grown up dramatically since z = 3 \citep{Daddi05,Trujillo07}. Detailed theoretical studies have shown that Es have undergone a two-phase formation process \citep{Oser10}. 
In the first phase, an initial rapid dissipative gas collapse produces a compact E which primarily hosts in-situ stars \citep{Khochfar06,Barro13,Wellons16}.
In the second, the growth mainly occurs through the accretion of satellite galaxies and the accreted satellite galaxies deposit their stellar population in the outskirt of the primary host, eventually making the host larger in overall size \citep{Naab09}.
The fraction of ex-situ stars is as high as 80\% in the case of high mass Es. 
It is well established that Es are the products of merged disk galaxies \citep{Springel05,Naab07,Duc15}, and galaxy mergers and accretions are fundamental to shape the observable properties of massive galaxies. 
On the other hand, the lower mass systems, a.k.a., dwarf galaxies, are predominantly made up of in-situ stars \citep{Cooper13,Pillepich18,Davison20}.
Thus, the evolution of dwarf galaxies is expected to be influenced by the environment-related mechanism more than the merger or accretion \citep{Boselli06,Kormendy09,Lisker09,Paudel14a}. 

\begin{figure}
\includegraphics[width=8.5cm]{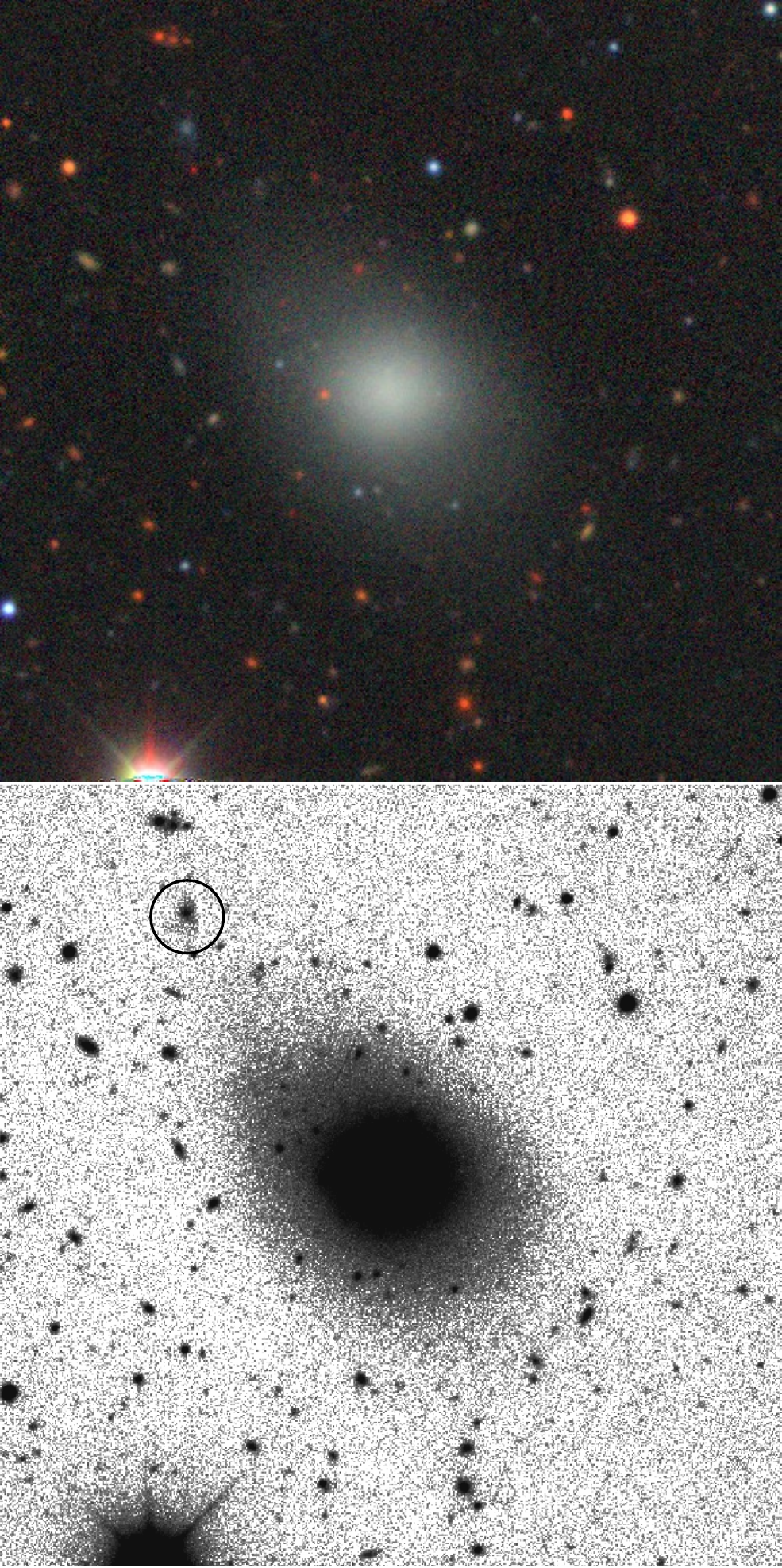}
\caption{dE1256 seen from the Legacy survey images. We show a $g-r-z$ combined color image in the top panel, which we obtained from the Legacy survey viewer tool. The bottom panel gray scale image is produced from co-adding $g$-$r$-$z$ band images which have a field of view of 2.5$\arcmin\times$2.5\arcmin. The color stretching is arbitrarily chosen to make the best view of low-surface brightness features. We identify a possible disrupted companion by a black circle of 7$\arcsec$ radius.}
\label{main}
\end{figure}

Large-scale observational studies reported that early-type dwarf galaxies (dEs) are primarily found in cluster and group environments, supporting their origin as the product of the environment \citep{Binggeli87,Geha12}. However, several recent deep imaging surveys revealed that some dEs might have experienced recent merger or accretion \cite{Paudel17,Paudel18}.

This work presents a detailed morphological study on a dE, SDSS\,J125651.47+163024.2 (hereafter dE1256), located in an almost isolated environment. dE1256 possesses a signature of ongoing interaction with an even smaller satellite galaxy whose stellar mass was once five times lower than that of dE1256. dE1256 has a luminosity M$_{g}$ = $-$14.88 mag, which is similar to the luminosity of previously discovered shell-feature merging dwarfs (VCC\,1361 and VCC\,1668) in the Virgo cluster \cite[hereafter P17]{Paudel17}.

\begin{figure*}
\includegraphics[width=18cm]{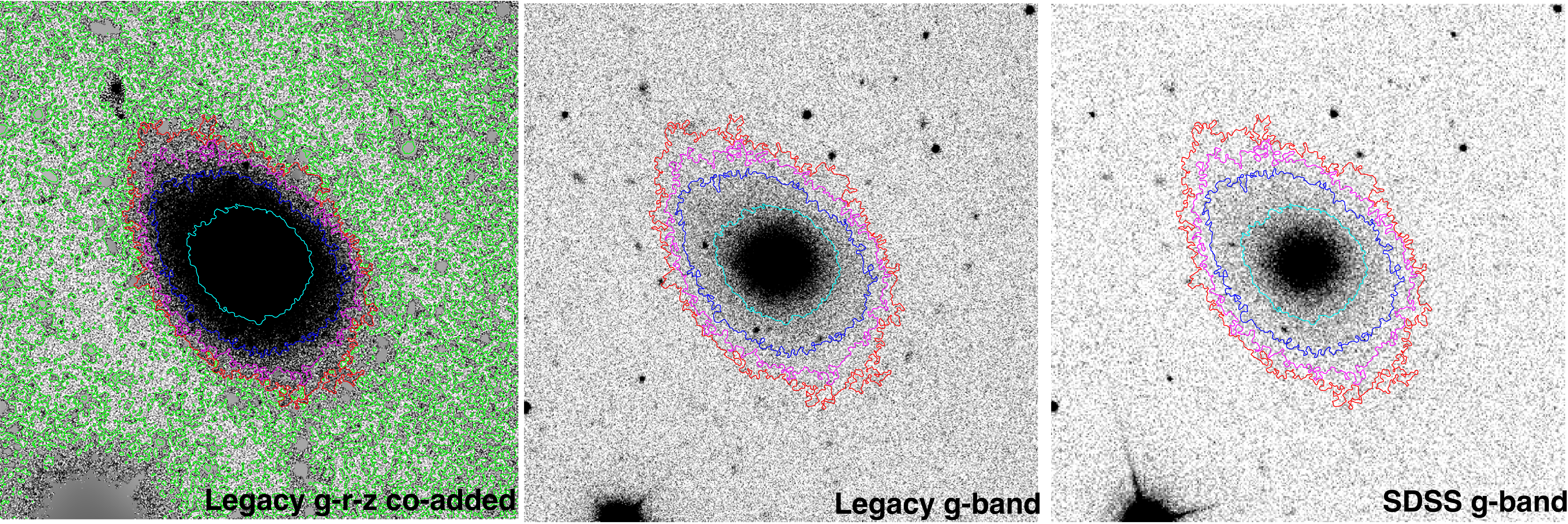}
\caption{We show a co-added $g-r-z$ high S/N image in the left panel, where the color (green, red, magenta, blue and cyan) contours represent surface brightness levels of 28, 27, 26, 25 and 24 mag arcsec$^{-2}$, respectively. In this image, all unrelated foreground and background sources are masked. 
To make a comparison between Legacy and SDSS image, we show their $g$-band image in the middle and right panel, respectively. We also show a color contour as a visual guidance, which was prepared in co-added image in the leftmost panel . It is clear that the Legacy survey image is deeper than the SDSS where the low surface brightness component is prominent.}
\label{comp}
\end{figure*}

\section{Environment and identification}
\begin{figure}
\includegraphics[width=8cm]{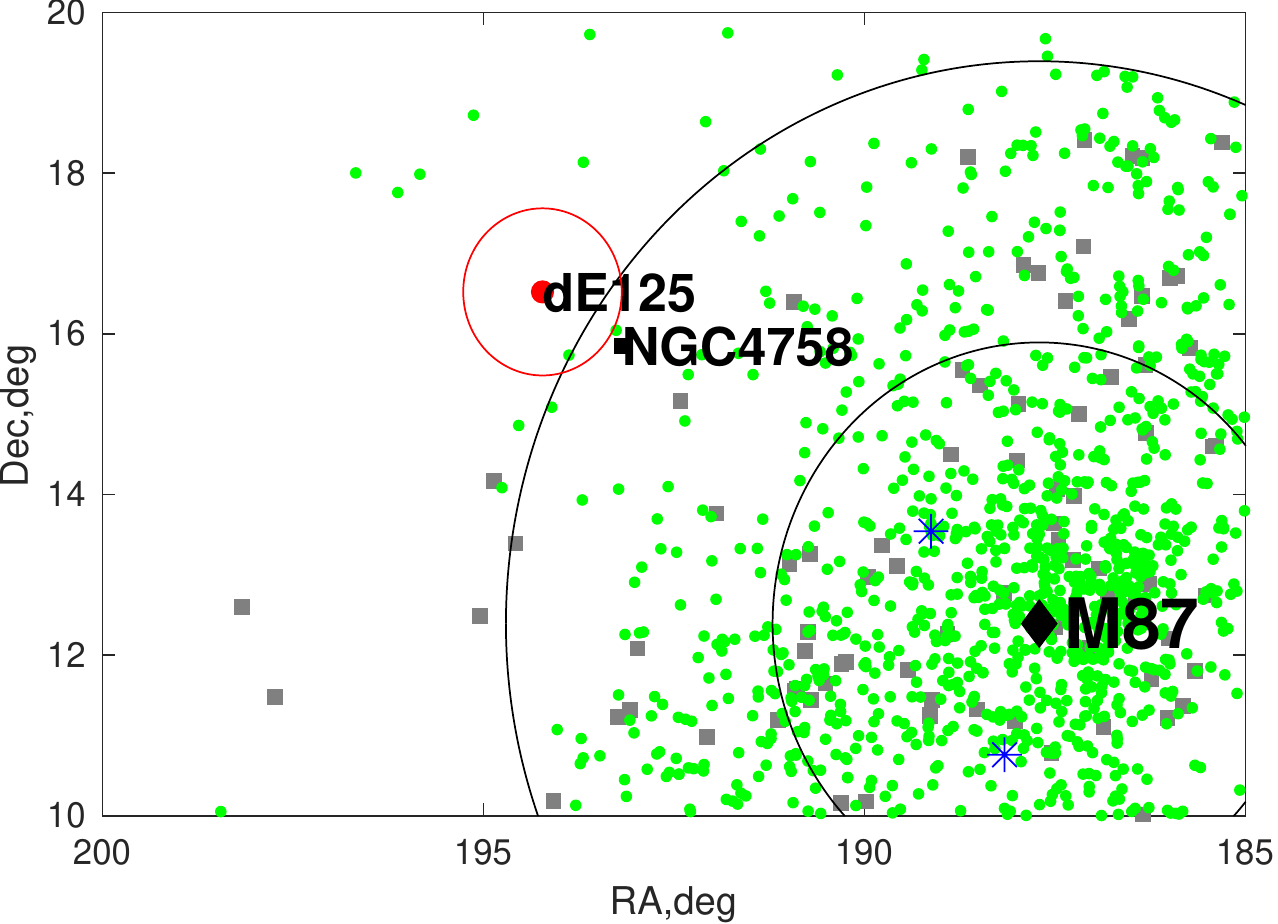}
\caption{Location of dE1256, next to the Virgo cluster. Green dots and gray squares represent dEs and bright galaxies, respectively. The centre of the Virgo cluster, position of M87, is marked with a large diamond symbol and the two black concentric rings represent sky-projected radii of 1 and 2 Mpc from M87. We highlight an area of 300 kpc radius around dE1256 with a red circle. The position of NGC\,4758 is marked by a black square symbol. With two blue asterisks, we also mark the positions of the Virgo cluster's shell-featured dEs that was studied in P17.
}
\label{asky}
\end{figure}

We have conducted a systematic search of low-mass early-type galaxies in diverse environments (clusters, groups, and fields) of the local volume (z $<$ 0.01). For this purpose, we visually inspected a large sky area observed by the SDSS and Legacy imaging surveys \citep{Aihara11,Dey19}, and identified 5054 dEs that are located in clusters, groups, and fields. 
The resultant main catalog will be published in Paudel et al. (2022, {\it submitted}). The strength of the catalog lies in detailing the morphological characteristics like tidal, merger, diffuse, and nucleated or non-nucleated features.
We found that a very small fraction ($\sim$0.3\%) of dEs possess the morphological features that could possibly originate from a merger. Among them, dE1256 is unique, in that it shows signs of ongoing accretion of a smaller dwarf galaxy (Figure \ref{main}).

dE1256 is located at z = 0.004 in the northeast of the Virgo cluster at an angular distance of $\sim$8$^{\circ}$ from M87, and has no significant massive neighbor (Figure \ref{asky}). 
Based on the NED, the nearest companion galaxy is NGC\,4758, which is located at a sky-projected distance of 370 kpc from the center of dE1256 in the southwest direction. NGC\,4758 is a star-forming low-mass galaxy of M$_{B}$ = $-$17.64 mag and has a radial velocity similar to that of dE1256, i.e., 1245 km s$^{-1}$. According to the main catalog, dE1256 is a member of the Virgo cluster and has an assigned distance of 16.5 Mpc. It is a non-nucleated galaxy with a mean surface brightness of 22.3 mag arcsec$^{-2}$.


\section{Data Analysis}\label{anal}

A view of dE1256 in the optical band images is shown in Figure \ref{main}. We show a $g-r-z$ combined color image in the top panel, which we obtained from the Legacy survey viewer tool. To get a higher signal of the low-surface brightness component, we co-added the legacy $g$, $r$ and $z$ band images, and show the resulting image in the lower panel. The extension of low-surface brightness component toward upper north-east direction is clear. After a careful inspection of the asymmetric tidal feature, we spot a possible disrupted companion located at an angular distance of 1$\arcmin$ (black circle) from the center of dE1256. 

In Figure \ref{comp}, we show the comparative picture of the Legacy (middle) and SDSS (right ) images. To make a quantitate comparison, we also show a co-added high-S/N image as a reference in left-most panel,  where we overlaid  color (green, red, magenta, blue and cyan) contours representing surface brightness levels of 28, 27, 26, 25, and 24 mag arcsec$^{-2}$, respectively. It is clear that the low surface brightness component can be detected as low as 27 mag arcsec$^{-2}$ level in the co-added image. While looking single filter $g$-band image in the middle and right panels, the legacy image is comparatively deeper than the SDSS and indeed, the extended tidal feature is much more evident in the Legacy image. However,  the similar appearance of the low surface brightness feature around dE1256 in the two images taken by two independent surveys, i.e., the SDSS and the Legacy, confirms that the observed low-surface brightness asymmetric tidal tail is not an artifact of image processing or instrumental defect.


In order to perform a detailed analysis of the morphology and structure of dE1256, we used the SDSS-III images \citep{Aihara11}. Although the Legacy images are deeper than the SDSS, we noticed that the sky-background subtraction is not as perfect as the SDSS. In the stacked and mosaiced Legacy images, the sky background seems over-subtracted, which creates an artificial decrease of the count at the outer parts of galaxies. The effect is severe for extended low-surface brightness objects. Therefore, we only used the SDSS images for the surface photometry and derived structural parameters. 

With help of the IRAF $ellipse$ task \citep{Jedrzejewski87}, we performed surface photometry on the $r$-band image of dE1256. Before running the ellipse task, we subtracted the sky-background as described in \cite{Paudel15} and manually masked background and foreground unrelated objects. Fitting ellipses to the isophotes of the surface brightnesses, we extracted azimuthally averaged radial surface brightness profiles along the major axis. In doing so, the center and position angle of the ellipse were held fixed. The ellipticity was allowed to vary. The galaxy center was calculated using the task $imcntr$, and input ellipse parameters were determined applying several iterative runs of the $ellipse$ task.

\begin{figure}
\includegraphics[width=8.5cm]{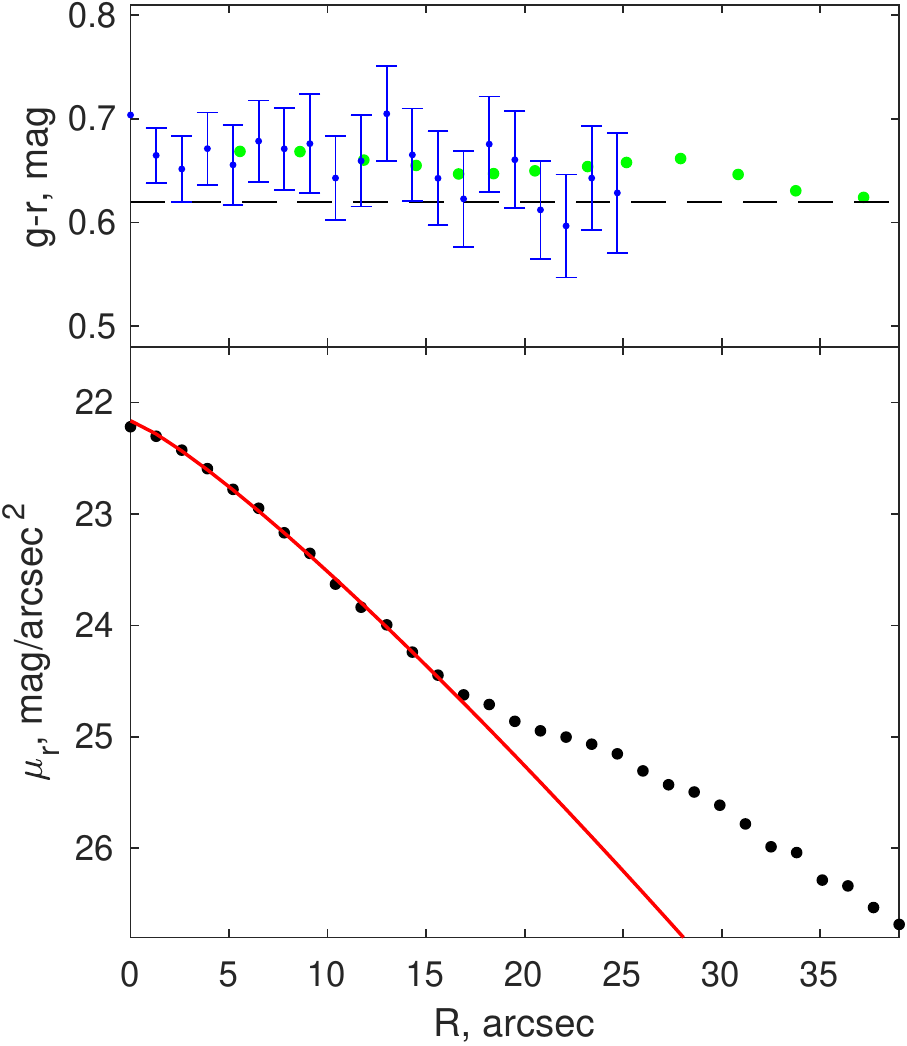}
\caption{(Top panel) The $g-r$ color profile of dE1256 along its major axis (blue dots with error bars).
We also derived the color profile (green dots) by selecting 5$\arcsec$\,$\times$\,5$\arcsec$ box region along the major axis. The estimated error-bar on these green points is in order of 0.02 mag which is significantly low compared to that of the blue dots. 
The black dash horizontal line represents the overall $g-r$ color of the putative disrupted companion. (Bottom) The SDSS $r$-band surface brightness profile of dE1256 along its major axis. 
We show best fitted S\'ersic profile for R $\leq$ 15$\arcsec$ in red (n = 0.63 and R$_{h}$ = 0.6 kpc).
}
\label{profile}
\end{figure}

In the lower panel of Figure \ref{profile}, we present the $r$-band major axis light profile of dE1256, which clearly shows a break at $\sim$15\arcsec (or $\mu_{r}$ = 24.7 mag arcsec$^{-2}$) attributing to the presence of a  dominant tidal tail at that region. We modeled the observed $r$-band major axis light profile with a S\'ersic function \citep{Sersic63}. Since the asymmetrical tidal feature dominates the outer part of dE1256, we skipped this region during the modeling. We only used the inner part (before the break) using a surface brightness cut of $\mu_{r}$ $<$ 23.7 mag arcsec$^{-2}$. 
With the help of $\chi^{2}$-minimization scheme, we obtained a best fit S\'ersic function with an index n = 0.63 and a half-light radius R$_{h}$ = 0.6 kpc. These values are typically representative of normal dE of luminosity M$_{g}$ $\simeq$ $-$13.5 mag \citep{Janz14}.

We show the $g-r$ color profile along the major axis in the upper panel of Figure \ref{profile}. The color profile was derived from the azimuthally averaged light profiles obtained from ellipse fitting in the $g$ and $r$ bands. To increase the signal-to-noise ratio in derived color indices, we also calculated the color profile along the major axis by selecting the 5$\arcsec$\,$\times$\,5$\arcsec$ box region along the major axis shown by green dots. The horizontal dash line represents the color index of the putative disrupted companion. We find that $g-r$ of dE1256, in general, declines at large galactocentric radii, and it matches fairly well with the color of the disruptive companion.

From a morphological perspective, dE1256 is likely undergoing a minor merger event. Although it is not trivial to calculate the merger ratio from the remnant, we used a two-component model---one is an undisturbed host at the center and the other is a destroyed tidal tail---to calculate the approximate stellar mass in each component. 
The supposed primary galaxy, which has an un-disturbed inner component, is modeled with a S\'ersic function, and thus the best-fitted inner component model represents the host galaxy. And then the leftover tidal feature (residual) is assumed to be entirely built up by an accreted minor galaxy.
The flux ratio between the two components is 0.2, and considering their similar stellar population properties suggested by their similar $g-r$ color, the flux ratio directly corresponds to stellar mass ratio. However, there is a reasonable chance that both components' stellar populations mix at any galactocentric distance. We might have underestimated the contribution of dE1256's stellar population as we used only its inner part to calculate the mass.

Because of the complexity in the observed light profile of dE1256, no overall structural parameters can be derived using a simple S\'ersic function modeling. To measure overall photometric properties, we used a non-parametric approach, i.e., the Petrosian method \citep[J08]{Janz08}. Using the Petrosian method, the estimated value of total luminosity and half-light radius are M$_{g}$ = $-$14.88 mag and R$_{h}$ = 1.2 kpc, respectively. It has $g-r$ color index of 0.66 mag, a typical color index of dEs of M$_{g}$ $\simeq$ $-$14.88 mag \citep{Lisker08,Janz09}. The survived part of the possible companion has a total luminosity of M$_{g}$ = $-$8.94 mag, and it has a $g-r$ color of 0.61 mag. We calculate the stellar masses using a calibration provided by \cite{Zhang17} for the $g-r$ color.  The calculated stellar masses of dE1256 and the companion are 1.3$\times$10$^{8}$ and 4.8$\times$10$^{5}$ M$_{\sun}$, respectively.

\begin{figure}
\includegraphics[width=8cm]{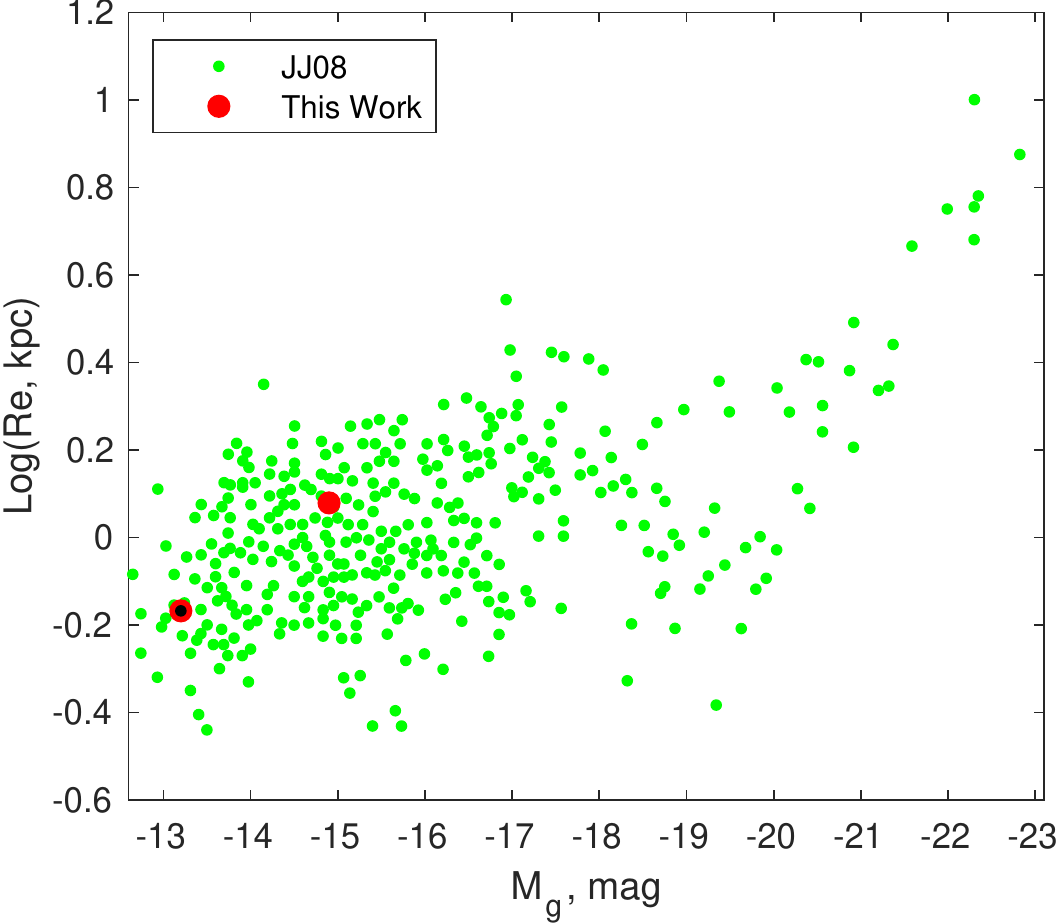}
\caption{The size$-$magnitude relation of early-type galaxies.   
The large red symbol represents the entire body of dE1256 and the same symbol with a black dot is for its inner S\'ersic-profiled component.
The comparison sample (small green symbols) is taken from J08.
}
\label{size}
\end{figure}

We find that the SDSS targeted the host galaxy dE1256 for spectroscopic observation. As expected, the SDSS fiber spectrum shows no emission in the Balmer lines. We derived simple stellar population (SSP) properties by fitting the SDSS spectrum with an SSP model. To estimate the SSP parameters, we followed a similar procedure that we employed in \cite{Paudel14b}. The procedure uses a publicly available full-spectrum fitting tool ULySS \citep{Koleva09}, which compares the observed spectrum with the SSP models of \cite{Vazdekis10}. 
The SSP age and metallicity of dE1256, derived based on the SDSS fiber spectroscopy, is 6.2$\pm$1.2 Gyr and log(Z/Z$_{\sun}$) = $-$0.26$\pm$0.07 dex, respectively.

Indeed, the SDSS fiber spectroscopy observes only the central 3$\arcsec$ region of a galaxy, and therefore, the derived stellar population properties pertain to the inner component. The typical survival time of a tidal tail is at most a couple of Giga years \citep{Mancillas19}, and a significantly larger SSP age than the tidal tail survival time suggests that dE1256 was already quenched before the interaction.

\section{ Summary and Discussion}\label{sec:dis}

We have presented a case of an interacting dwarf galaxy system in a nearly isolated environment. From our primary catalog of 5405 dEs, there are only 13 dEs that are classified as merger remnants, and among them, only dE1256 exhibits a feature other than the shell. \cite{Paudel17} have shown that symmetric shell features are more likely a product of an equal mass merger with a head-on collision, where the stellar populations of both interacting galaxies can redistribute homogeneously at any galactocentric radii. Indeed, the tidal tail in dE1256 tells a different story presenting a unique perspective on growth of a low-mass galaxy with the minor merger, which mainly deposits the stellar populations at the outer parts. The growth of galaxies through  hierarchical merging is a natural phenomenon of our well-known $\Lambda$CDM cosmology \citep[e.g.,][]{Springel05}. However, at the lower mass scale, it has been shown through observation and theory, that the galaxy-galaxy merger is a rare phenomenon \citep[e.g.,][]{Lucia06}.

\begin{figure}
\includegraphics[width=8.5cm]{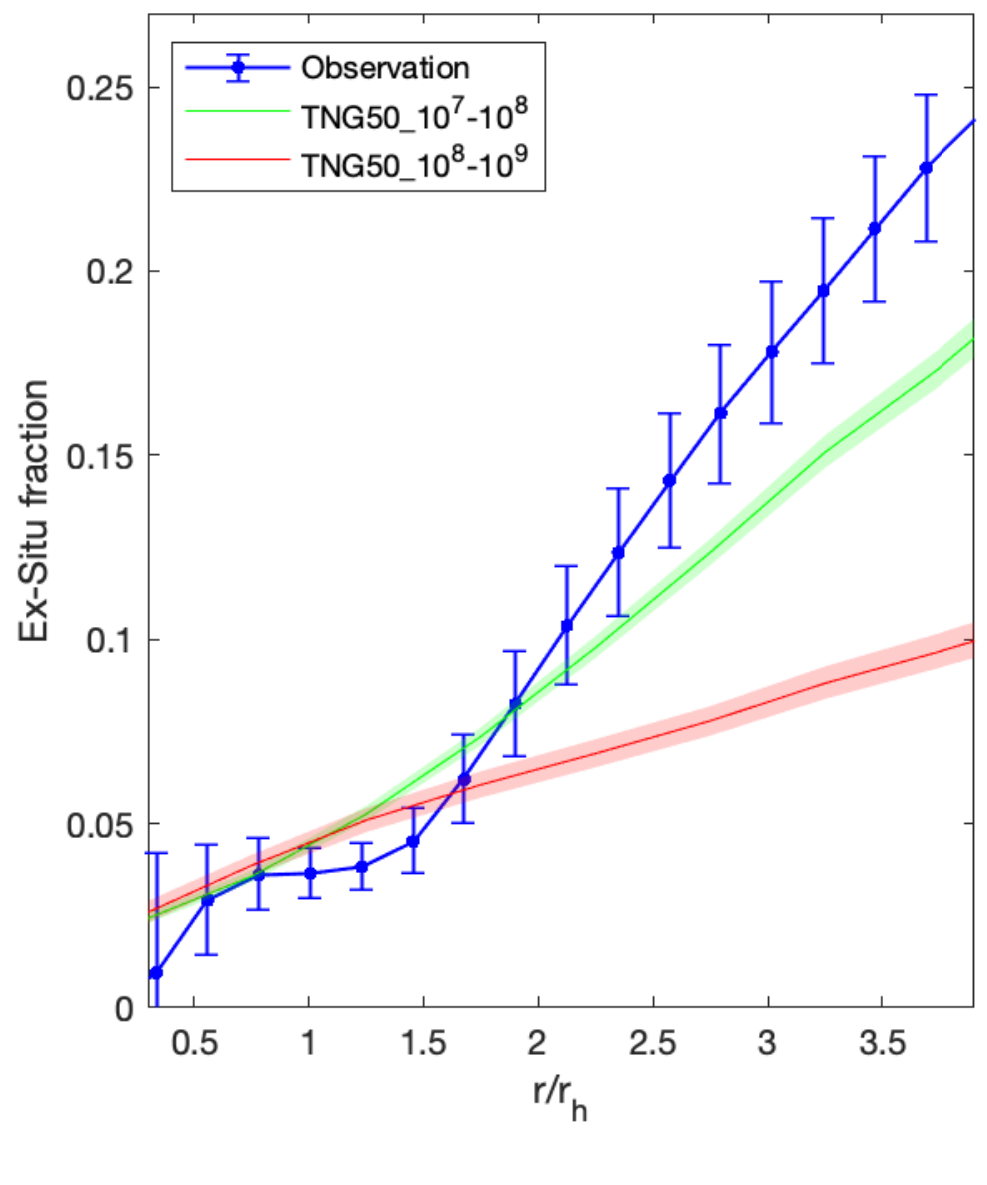}
\caption{The radial profile of the ex-situ stellar mass fraction for early-type galaxies. The blue line represents dE1256. The errorbar represents uncertainty in photometric measurement which does not include the uncertainty in the S\'ersic modeling. We show the result of TNG50 simulation in red and green, which represent binned sample of M$_{stellar}$ = 10$^{7}$\,$\sim$\,10$^{8}$ M$_{\sun}$ (7047 galaxies) and M$_{stellar}$ = 10$^{8}$\,$\sim$\,10$^{9}$ M$_{\sun}$ (2546 galaxies), respectively. The shade regions show the 95 percent confidence intervals estimated by 1000 bootstrap resamplings of the simulated galaxies. See section \ref{sec:dis} for the detail.
}
\label{compare}
\end{figure}

\cite{Paudel18} presented an extensive survey of merging dwarf galaxies in the local universe and showed that the dry-merger between dwarf galaxies is a rare phenomenon. There are only three early-type merging dwarfs out of 177, and dE1256 is one of them. On the other hand, there are three shell feature dEs in the Virgo cluster and dE1256 is the fourth dE that possesses the tidal feature originating from the dwarf-dwarf merger.

To quantify the merger parameter --namely merger ratio, we have used a simple approach fitting the galaxy light profile into two components model --one inner primary and another outer secondary. We assume that the outer secondary component is entirely built up by the accretion of another dwarf. We know that dEs possess multi-component surface brightness profile \citep{Jerjen20,Janz14,Lee18}, and there are as high as 60\% dEs that has such multi-component features at the brighter end of their population \citep{Lisker07}. The origin of such a multi-component light profile, in general, is attributed to the presence of a disk feature, like a bar or a spiral arm, an indication of the environmental transformation of a late-type galaxy to dE through the mechanism such as ram-pressure stripping or tidal harassment \citep{Boselli06,Smith21}.
dE1256 inner core looks rather regular and non-nucleated, and the outer extended tidal tail is asymmetric. Therefore, we have interpreted that the observed multi-component light profile with outer tidal features originated from the tidal disruption of a smaller dwarf galaxy.

In Figure \ref{size}, we show the size$-$magnitude relation of dEs, where the comparison sample is taken from \cite{Janz08}. We see two different regimes of the relation; in the bright part, the relation is steep, and in the faint part, the relation is nearly constant. As identified by the red circle, dE1256 falls in the faint regime, where the data point shows a large scatter. The red circle with a black dot represents an inner component that is significantly smaller than the overall size of dE1256 (shown as the red circle). We find that while the accretion contributes only 20\% of the overall stellar population, the size of dE1256 increases by a factor of two after the accretion.

Figure \ref{compare} displays the derived ex-situ stellar mass fraction of dE1256 as a function of galactocentric radius. For the comparison, we overplot the radial ex-situ fraction derived from the TNG50 simulation of the IllustrisTNG project \citep{Nelson19a,Nelson19b,Pillepich19}. For this, we produced stacked profiles of the ex-situ mass fraction of all early-type subhalos in stellar mass bins of M$_{stellar}$ = 10$^{7}$\,$\sim$\,10$^{8}$ M$_{\sun}$ and M$_{stellar}$ = 10$^{8}$\,$\sim$\,10$^{9}$ M$_{\sun}$, respectively. The subhalos are classified as early-type subhalos when their specific star-formation rate (sSFR) is as low as log(sSFR/yr$^{-1}$)\,$<$\,$-$10.
Ex-situ stars were defined as those formed outside the main progenitor, following the method of \cite{Rodriguez16} and \cite{Davison20}. 
We discarded a small fraction of stellar particles formed at very early times when the main progenitor has not yet been detected. 
We find that our estimated ex-situ stellar-mass profile for dE1256 agrees, in general, with the simulated galaxy ex-situ stellar mass profiles. 
Compared to the TNG50 result, it seems that we have slightly underestimated the ex-situ stellar mass fractions at the inner region and overestimated at the outer region. 
Such deviation can be a consequence of our simple assumption that the host dominates the inner component and the disrupted dwarf satellite primarily deposits its stellar population at the outer radii. Moreover, the simulation estimates the ex-situ fraction based on the actual merging history, yet in observation, we use a simple assumption that the observed dual-component light profile is due to the different stellar populations of the host and satellite. Compared to the simulation result, the high ex-situ fraction at the outer region of dE1256 is atypical, pointing to an ongoing tidal accretion of a satellite.

\section*{Acknowledgements}
S.P. and S.-J.Y., respectively, acknowledge support from the New Researcher Program (Shinjin grant No. 2019R1C1C1009600) and the Midcareer Researcher Program (No. 2019R1A2C3006242) through the National Research Foundation of Korea.

This study is based on archival images and spectra from the Sloan Digital Sky Survey and images from the Legacy survey. The full acknowledgment for the SDSS can be found at http://www.sdss.org/collaboration/credits.html.  Funding for the SDSS has been provided by the Alfred P. Sloan Foundation, the Participating Institutions, the National Science Foundation, the U.S. Department of Energy, the National Aeronautics and Space Administration, the Japanese Monbukagakusho, the Max Planck Society, and the Higher Education Funding Council for England. The SDSS website is http://www.sdss.org/. 

The Legacy Surveys consist of three individual and complementary projects: the Dark Energy Camera Legacy Survey (DECaLS; NOAO Proposal ID \# 2014B-0404; PIs: David Schlegel and Arjun Dey), the Beijing-Arizona Sky Survey (BASS; NOAO Proposal ID \# 2015A-0801; PIs: Zhou Xu and Xiaohui Fan), and the Mayall $z$-band Legacy Survey (MzLS; NOAO Proposal ID \# 2016A-0453; PI: Arjun Dey). DECaLS, BASS, and MzLS together include data obtained, respectively, at the Blanco telescope, Cerro Tololo Inter-American Observatory, National Optical Astronomy Observatory (NOAO); the Bok telescope, Steward Observatory, University of Arizona; and the Mayall telescope, Kitt Peak National Observatory, NOAO. The Legacy Surveys project is honored to be permitted to conduct astronomical research on Iolkam Du'ag (Kitt Peak), a mountain with particular significance to the Tohono O'odham Nation. The full acknowledgment for the Legacy Surveys can be found at https://www.legacysurvey.org/acknowledgment

The IllustrisTNG simulations were undertaken with compute time awarded by the Gauss Centre for Supercomputing (GCS) under GCS Large-Scale Projects GCS-ILLU and GCS-DWAR on the GCS share of the supercomputer Hazel Hen at the High Performance Computing Center Stuttgart (HLRS), as well as on the machines of the Max Planck Computing and Data Facility (MPCDF) in Garching, Germany

\section{DATA AVAILABILITY}
Most of the data underlying this article are publicly available.
The derived data generated in this research will also be shared on reasonable request to the corresponding author.



\bsp	
\label{lastpage}

\begin{thebibliography}{}
\makeatletter
\relax
\def\mn@urlcharsother{\let\do\@makeother \do\$\do\&\do\#\do\^\do\_\do\%\do\~}
\def\mn@doi{\begingroup\mn@urlcharsother \@ifnextchar [ {\mn@doi@}
  {\mn@doi@[]}}
\def\mn@doi@[#1]#2{\def\@tempa{#1}\ifx\@tempa\@empty \href
  {http://dx.doi.org/#2} {doi:#2}\else \href {http://dx.doi.org/#2} {#1}\fi
  \endgroup}
\def\mn@eprint#1#2{\mn@eprint@#1:#2::\@nil}
\def\mn@eprint@arXiv#1{\href {http://arxiv.org/abs/#1} {{\tt arXiv:#1}}}
\def\mn@eprint@dblp#1{\href {http://dblp.uni-trier.de/rec/bibtex/#1.xml}
  {dblp:#1}}
\def\mn@eprint@#1:#2:#3:#4\@nil{\def\@tempa {#1}\def\@tempb {#2}\def\@tempc
  {#3}\ifx \@tempc \@empty \let \@tempc \@tempb \let \@tempb \@tempa \fi \ifx
  \@tempb \@empty \def\@tempb {arXiv}\fi \@ifundefined
  {mn@eprint@\@tempb}{\@tempb:\@tempc}{\expandafter \expandafter \csname
  mn@eprint@\@tempb\endcsname \expandafter{\@tempc}}}

\bibitem[\protect\citeauthoryear{{Aihara} et~al.,}{{Aihara}
  et~al.}{2011}]{Aihara11}
{Aihara} H.,  et~al., 2011, \mn@doi [\apjs] {10.1088/0067-0049/193/2/29}, \href
  {http://adsabs.harvard.edu/abs/2011ApJS..193...29A} {193, 29}

\bibitem[\protect\citeauthoryear{{Barro} et~al.,}{{Barro}
  et~al.}{2013}]{Barro13}
{Barro} G.,  et~al., 2013, \mn@doi [\apj] {10.1088/0004-637X/765/2/104}, \href
  {https://ui.adsabs.harvard.edu/abs/2013ApJ...765..104B} {765, 104}

\bibitem[\protect\citeauthoryear{{Binggeli}, {Tammann}  \&
  {Sandage}}{{Binggeli} et~al.}{1987}]{Binggeli87}
{Binggeli} B.,  {Tammann} G.~A.,   {Sandage} A.,  1987, \mn@doi [\aj]
  {10.1086/114467}, \href
  {https://ui.adsabs.harvard.edu/abs/1987AJ.....94..251B} {94, 251}

\bibitem[\protect\citeauthoryear{{Boselli} \& {Gavazzi}}{{Boselli} \&
  {Gavazzi}}{2006}]{Boselli06}
{Boselli} A.,  {Gavazzi} G.,  2006, \mn@doi [\pasp] {10.1086/500691}, \href
  {https://ui.adsabs.harvard.edu/abs/2006PASP..118..517B} {118, 517}

\bibitem[\protect\citeauthoryear{{Cooper}, {D'Souza}, {Kauffmann}, {Wang},
  {Boylan-Kolchin}, {Guo}, {Frenk}  \& {White}}{{Cooper}
  et~al.}{2013}]{Cooper13}
{Cooper} A.~P.,  {D'Souza} R.,  {Kauffmann} G.,  {Wang} J.,  {Boylan-Kolchin}
  M.,  {Guo} Q.,  {Frenk} C.~S.,   {White} S. D.~M.,  2013, \mn@doi [\mnras]
  {10.1093/mnras/stt1245}, \href
  {https://ui.adsabs.harvard.edu/abs/2013MNRAS.434.3348C} {434, 3348}

\bibitem[\protect\citeauthoryear{{Daddi} et~al.,}{{Daddi}
  et~al.}{2005}]{Daddi05}
{Daddi} E.,  et~al., 2005, \mn@doi [\apj] {10.1086/430104}, \href
  {https://ui.adsabs.harvard.edu/abs/2005ApJ...626..680D} {626, 680}

\bibitem[\protect\citeauthoryear{{Davison}, {Norris}, {Pfeffer}, {Davies}  \&
  {Crain}}{{Davison} et~al.}{2020}]{Davison20}
{Davison} T.~A.,  {Norris} M.~A.,  {Pfeffer} J.~L.,  {Davies} J.~J.,   {Crain}
  R.~A.,  2020, \mn@doi [\mnras] {10.1093/mnras/staa1816}, \href
  {https://ui.adsabs.harvard.edu/abs/2020MNRAS.497...81D} {497, 81}

\bibitem[\protect\citeauthoryear{{De Lucia}, {Springel}, {White}, {Croton}  \&
  {Kauffmann}}{{De Lucia} et~al.}{2006}]{Lucia06}
{De Lucia} G.,  {Springel} V.,  {White} S. D.~M.,  {Croton} D.,   {Kauffmann}
  G.,  2006, \mn@doi [\mnras] {10.1111/j.1365-2966.2005.09879.x}, \href
  {https://ui.adsabs.harvard.edu/abs/2006MNRAS.366..499D} {366, 499}

\bibitem[\protect\citeauthoryear{{Dey} et~al.,}{{Dey} et~al.}{2019}]{Dey19}
{Dey} A.,  et~al., 2019, \mn@doi [\aj] {10.3847/1538-3881/ab089d}, \href
  {https://ui.adsabs.harvard.edu/abs/2019AJ....157..168D} {157, 168}

\bibitem[\protect\citeauthoryear{{Duc} et~al.,}{{Duc} et~al.}{2015}]{Duc15}
{Duc} P.-A.,  et~al., 2015, \mn@doi [\mnras] {10.1093/mnras/stu2019}, \href
  {https://ui.adsabs.harvard.edu/abs/2015MNRAS.446..120D} {446, 120}

\bibitem[\protect\citeauthoryear{{Geha}, {Blanton}, {Yan}  \& {Tinker}}{{Geha}
  et~al.}{2012}]{Geha12}
{Geha} M.,  {Blanton} M.~R.,  {Yan} R.,   {Tinker} J.~L.,  2012, \mn@doi [\apj]
  {10.1088/0004-637X/757/1/85}, \href
  {https://ui.adsabs.harvard.edu/abs/2012ApJ...757...85G} {757, 85}

\bibitem[\protect\citeauthoryear{{Janz} \& {Lisker}}{{Janz} \&
  {Lisker}}{2008}]{Janz08}
{Janz} J.,  {Lisker} T.,  2008, \mn@doi [\apjl] {10.1086/595720}, \href
  {https://ui.adsabs.harvard.edu/abs/2008ApJ...689L..25J} {689, L25}

\bibitem[\protect\citeauthoryear{{Janz} \& {Lisker}}{{Janz} \&
  {Lisker}}{2009}]{Janz09}
{Janz} J.,  {Lisker} T.,  2009, \mn@doi [\apjl] {10.1088/0004-637X/696/1/L102},
  \href {https://ui.adsabs.harvard.edu/abs/2009ApJ...696L.102J} {696, L102}

\bibitem[\protect\citeauthoryear{{Janz} et~al.,}{{Janz} et~al.}{2014}]{Janz14}
{Janz} J.,  et~al., 2014, \mn@doi [\apj] {10.1088/0004-637X/786/2/105}, \href
  {https://ui.adsabs.harvard.edu/abs/2014ApJ...786..105J} {786, 105}

\bibitem[\protect\citeauthoryear{{Jedrzejewski}}{{Jedrzejewski}}{1987}]{Jedrzejewski87}
{Jedrzejewski} R.~I.,  1987, \mn@doi [\mnras] {10.1093/mnras/226.4.747}, \href
  {https://ui.adsabs.harvard.edu/abs/1987MNRAS.226..747J} {226, 747}

\bibitem[\protect\citeauthoryear{{Jerjen}, {Kalnajs}  \& {Binggeli}}{{Jerjen}
  et~al.}{2000}]{Jerjen20}
{Jerjen} H.,  {Kalnajs} A.,   {Binggeli} B.,  2000, \aap, \href
  {https://ui.adsabs.harvard.edu/abs/2000A&A...358..845J} {358, 845}

\bibitem[\protect\citeauthoryear{{Khochfar} \& {Silk}}{{Khochfar} \&
  {Silk}}{2006}]{Khochfar06}
{Khochfar} S.,  {Silk} J.,  2006, \mn@doi [\apjl] {10.1086/507768}, \href
  {https://ui.adsabs.harvard.edu/abs/2006ApJ...648L..21K} {648, L21}

\bibitem[\protect\citeauthoryear{{Koleva}, {Prugniel}, {Bouchard}  \&
  {Wu}}{{Koleva} et~al.}{2009}]{Koleva09}
{Koleva} M.,  {Prugniel} P.,  {Bouchard} A.,   {Wu} Y.,  2009, \mn@doi [\aap]
  {10.1051/0004-6361/200811467}, \href
  {https://ui.adsabs.harvard.edu/abs/2009A&A...501.1269K} {501, 1269}

\bibitem[\protect\citeauthoryear{{Kormendy}, {Fisher}, {Cornell}  \&
  {Bender}}{{Kormendy} et~al.}{2009}]{Kormendy09}
{Kormendy} J.,  {Fisher} D.~B.,  {Cornell} M.~E.,   {Bender} R.,  2009, \mn@doi
  [\apjs] {10.1088/0067-0049/182/1/216}, \href
  {https://ui.adsabs.harvard.edu/abs/2009ApJS..182..216K} {182, 216}

\bibitem[\protect\citeauthoryear{{Lee}, {Park}, {Kim}, {Moon}, {Lee}, {Kim}  \&
  {Cha}}{{Lee} et~al.}{2018}]{Lee18}
{Lee} Y.,  {Park} H.~S.,  {Kim} S.~C.,  {Moon} D.-S.,  {Lee} J.-J.,  {Kim}
  D.-J.,   {Cha} S.-M.,  2018, \mn@doi [\apj] {10.3847/1538-4357/aabc53}, \href
  {https://ui.adsabs.harvard.edu/abs/2018ApJ...859....5L} {859, 5}

\bibitem[\protect\citeauthoryear{{Lisker}}{{Lisker}}{2009}]{Lisker09}
{Lisker} T.,  2009, \mn@doi [Astronomische Nachrichten]
  {10.1002/asna.200911291}, \href
  {https://ui.adsabs.harvard.edu/abs/2009AN....330.1043L} {330, 1043}

\bibitem[\protect\citeauthoryear{{Lisker}, {Grebel}, {Binggeli}  \&
  {Glatt}}{{Lisker} et~al.}{2007}]{Lisker07}
{Lisker} T.,  {Grebel} E.~K.,  {Binggeli} B.,   {Glatt} K.,  2007, \mn@doi
  [\apj] {10.1086/513090}, \href
  {https://ui.adsabs.harvard.edu/abs/2007ApJ...660.1186L} {660, 1186}

\bibitem[\protect\citeauthoryear{{Lisker}, {Grebel}  \& {Binggeli}}{{Lisker}
  et~al.}{2008}]{Lisker08}
{Lisker} T.,  {Grebel} E.~K.,   {Binggeli} B.,  2008, \mn@doi [\aj]
  {10.1088/0004-6256/135/1/380}, \href
  {https://ui.adsabs.harvard.edu/abs/2008AJ....135..380L} {135, 380}

\bibitem[\protect\citeauthoryear{{Mancillas}, {Duc}, {Combes}, {Bournaud},
  {Emsellem}, {Martig}  \& {Michel-Dansac}}{{Mancillas}
  et~al.}{2019}]{Mancillas19}
{Mancillas} B.,  {Duc} P.-A.,  {Combes} F.,  {Bournaud} F.,  {Emsellem} E.,
  {Martig} M.,   {Michel-Dansac} L.,  2019, \mn@doi [\aap]
  {10.1051/0004-6361/201936320}, \href
  {https://ui.adsabs.harvard.edu/abs/2019A&A...632A.122M} {632, A122}

\bibitem[\protect\citeauthoryear{{Naab}, {Johansson}, {Ostriker}  \&
  {Efstathiou}}{{Naab} et~al.}{2007}]{Naab07}
{Naab} T.,  {Johansson} P.~H.,  {Ostriker} J.~P.,   {Efstathiou} G.,  2007,
  \mn@doi [\apj] {10.1086/510841}, \href
  {https://ui.adsabs.harvard.edu/abs/2007ApJ...658..710N} {658, 710}

\bibitem[\protect\citeauthoryear{{Naab}, {Johansson}  \& {Ostriker}}{{Naab}
  et~al.}{2009}]{Naab09}
{Naab} T.,  {Johansson} P.~H.,   {Ostriker} J.~P.,  2009, \mn@doi [\apjl]
  {10.1088/0004-637X/699/2/L178}, \href
  {https://ui.adsabs.harvard.edu/abs/2009ApJ...699L.178N} {699, L178}

\bibitem[\protect\citeauthoryear{{Nelson} et~al.,}{{Nelson}
  et~al.}{2019a}]{Nelson19a}
{Nelson} D.,  et~al., 2019a, \mn@doi [Computational Astrophysics and Cosmology]
  {10.1186/s40668-019-0028-x}, \href
  {https://ui.adsabs.harvard.edu/abs/2019ComAC...6....2N} {6, 2}

\bibitem[\protect\citeauthoryear{{Nelson} et~al.,}{{Nelson}
  et~al.}{2019b}]{Nelson19b}
{Nelson} D.,  et~al., 2019b, \mn@doi [\mnras] {10.1093/mnras/stz2306}, \href
  {https://ui.adsabs.harvard.edu/abs/2019MNRAS.490.3234N} {490, 3234}

\bibitem[\protect\citeauthoryear{{Oser}, {Ostriker}, {Naab}, {Johansson}  \&
  {Burkert}}{{Oser} et~al.}{2010}]{Oser10}
{Oser} L.,  {Ostriker} J.~P.,  {Naab} T.,  {Johansson} P.~H.,   {Burkert} A.,
  2010, \mn@doi [\apj] {10.1088/0004-637X/725/2/2312}, \href
  {https://ui.adsabs.harvard.edu/abs/2010ApJ...725.2312O} {725, 2312}

\bibitem[\protect\citeauthoryear{{Paudel} \& {Ree}}{{Paudel} \&
  {Ree}}{2014}]{Paudel14a}
{Paudel} S.,  {Ree} C.~H.,  2014, \mn@doi [\apjl]
  {10.1088/2041-8205/796/1/L14}, \href
  {http://adsabs.harvard.edu/abs/2014ApJ...796L..14P} {796, L14}

\bibitem[\protect\citeauthoryear{{Paudel}, {Lisker}, {Hansson}  \&
  {Huxor}}{{Paudel} et~al.}{2014}]{Paudel14b}
{Paudel} S.,  {Lisker} T.,  {Hansson} K.~S.~A.,   {Huxor} A.~P.,  2014, \mn@doi
  [\mnras] {10.1093/mnras/stu1171}, \href
  {http://adsabs.harvard.edu/abs/2014MNRAS.443..446P} {443, 446}

\bibitem[\protect\citeauthoryear{{Paudel}, {Duc}  \& {Ree}}{{Paudel}
  et~al.}{2015}]{Paudel15}
{Paudel} S.,  {Duc} P.~A.,   {Ree} C.~H.,  2015, \mn@doi [\aj]
  {10.1088/0004-6256/149/3/114}, \href
  {http://adsabs.harvard.edu/abs/2015AJ....149..114P} {149, 114}

\bibitem[\protect\citeauthoryear{{Paudel} et~al.,}{{Paudel}
  et~al.}{2017}]{Paudel17}
{Paudel} S.,  et~al., 2017, \mn@doi [\apj] {10.3847/1538-4357/834/1/66}, \href
  {http://adsabs.harvard.edu/abs/2017ApJ...834...66P} {834, 66}

\bibitem[\protect\citeauthoryear{{Paudel}, {Smith}, {Yoon},
  {Calder{\'o}n-Castillo}  \& {Duc}}{{Paudel} et~al.}{2018}]{Paudel18}
{Paudel} S.,  {Smith} R.,  {Yoon} S.~J.,  {Calder{\'o}n-Castillo} P.,   {Duc}
  P.-A.,  2018, \mn@doi [\apjs] {10.3847/1538-4365/aad555}, \href
  {https://ui.adsabs.harvard.edu/abs/2018ApJS..237...36P} {237, 36}

\bibitem[\protect\citeauthoryear{{Pillepich} et~al.,}{{Pillepich}
  et~al.}{2018}]{Pillepich18}
{Pillepich} A.,  et~al., 2018, \mn@doi [\mnras] {10.1093/mnras/stx3112}, \href
  {https://ui.adsabs.harvard.edu/abs/2018MNRAS.475..648P} {475, 648}

\bibitem[\protect\citeauthoryear{{Pillepich} et~al.,}{{Pillepich}
  et~al.}{2019}]{Pillepich19}
{Pillepich} A.,  et~al., 2019, \mn@doi [\mnras] {10.1093/mnras/stz2338}, \href
  {https://ui.adsabs.harvard.edu/abs/2019MNRAS.490.3196P} {490, 3196}

\bibitem[\protect\citeauthoryear{{Rodriguez-Gomez} et~al.,}{{Rodriguez-Gomez}
  et~al.}{2016}]{Rodriguez16}
{Rodriguez-Gomez} V.,  et~al., 2016, \mn@doi [\mnras] {10.1093/mnras/stw456},
  \href {https://ui.adsabs.harvard.edu/abs/2016MNRAS.458.2371R} {458, 2371}

\bibitem[\protect\citeauthoryear{{S{\'e}rsic}}{{S{\'e}rsic}}{1963}]{Sersic63}
{S{\'e}rsic} J.~L.,  1963, Boletin de la Asociacion Argentina de Astronomia La
  Plata Argentina, \href
  {https://ui.adsabs.harvard.edu/abs/1963BAAA....6...41S} {6, 41}

\bibitem[\protect\citeauthoryear{{Smith} et~al.,}{{Smith}
  et~al.}{2021}]{Smith21}
{Smith} R.,  et~al., 2021, \mn@doi [\apj] {10.3847/1538-4357/abe1b1}, \href
  {https://ui.adsabs.harvard.edu/abs/2021ApJ...912..149S} {912, 149}

\bibitem[\protect\citeauthoryear{{Springel} et~al.,}{{Springel}
  et~al.}{2005}]{Springel05}
{Springel} V.,  et~al., 2005, \mn@doi [\nat] {10.1038/nature03597}, \href
  {https://ui.adsabs.harvard.edu/abs/2005Natur.435..629S} {435, 629}

\bibitem[\protect\citeauthoryear{{Trujillo}, {Conselice}, {Bundy}, {Cooper},
  {Eisenhardt}  \& {Ellis}}{{Trujillo} et~al.}{2007}]{Trujillo07}
{Trujillo} I.,  {Conselice} C.~J.,  {Bundy} K.,  {Cooper} M.~C.,  {Eisenhardt}
  P.,   {Ellis} R.~S.,  2007, \mn@doi [\mnras]
  {10.1111/j.1365-2966.2007.12388.x}, \href
  {https://ui.adsabs.harvard.edu/abs/2007MNRAS.382..109T} {382, 109}

\bibitem[\protect\citeauthoryear{{Vazdekis}, {S{\'a}nchez-Bl{\'a}zquez},
  {Falc{\'o}n-Barroso}, {Cenarro}, {Beasley}, {Cardiel}, {Gorgas}  \&
  {Peletier}}{{Vazdekis} et~al.}{2010}]{Vazdekis10}
{Vazdekis} A.,  {S{\'a}nchez-Bl{\'a}zquez} P.,  {Falc{\'o}n-Barroso} J.,
  {Cenarro} A.~J.,  {Beasley} M.~A.,  {Cardiel} N.,  {Gorgas} J.,   {Peletier}
  R.~F.,  2010, \mn@doi [\mnras] {10.1111/j.1365-2966.2010.16407.x}, \href
  {https://ui.adsabs.harvard.edu/abs/2010MNRAS.404.1639V} {404, 1639}

\bibitem[\protect\citeauthoryear{{Wellons} et~al.,}{{Wellons}
  et~al.}{2016}]{Wellons16}
{Wellons} S.,  et~al., 2016, \mn@doi [\mnras] {10.1093/mnras/stv2738}, \href
  {https://ui.adsabs.harvard.edu/abs/2016MNRAS.456.1030W} {456, 1030}

\bibitem[\protect\citeauthoryear{{Zhang}, {Puzia}  \& {Weisz}}{{Zhang}
  et~al.}{2017}]{Zhang17}
{Zhang} H.-X.,  {Puzia} T.~H.,   {Weisz} D.~R.,  2017, \mn@doi [\apjs]
  {10.3847/1538-4365/aa937b}, \href
  {https://ui.adsabs.harvard.edu/abs/2017ApJS..233...13Z} {233, 13}

\makeatother
\end{thebibliography}
\end{document}